\documentclass[seceq]{ptptex}

\usepackage{graphicx}

\title{
$\Lambda (1405)$ from lattice QCD%
}

\author{
Toru T. \textsc{Takahashi}$^{a,b}$ and Makoto \textsc{Oka}$^c$%
}

\inst{
$^a$Yukawa Institute for Theoretical Physics,
Sakyo, Kyoto 606-8502, Japan\\
$^b$Gunma National College of Technology, Maebashi, Gunma
371-8530, Japan\\
$^c$Department of Physics, H-27, Tokyo Institute of Technology, Meguro, Tokyo 152-8551 Japan
}

\abst{
Low-lying $\Lambda$ baryons with spin 1/2 are 
analyzed in two-flavor lattice QCD.
In order to extract two low-lying states for each parity,
we construct $2 \times 2$ cross correlators 
from flavor SU(3) ``octet'' and ``singlet'' baryon operators,
and diagonalize them.
Two-flavor CP-PACS gauge configurations are employed, 
which are generated with the renormalization-group improved gauge action
and the ${\mathcal O}(a)$-improved quark action.
Simulation are performed at three different $\beta$'s, 
$\beta = 1.80$, 1.95 and 2.10,
whose corresponding lattice spacings are $a = 0.2150$, 0.1555 and 0.1076 fm. 
For each cutoff, we adopt four different hopping parameters,
($\kappa_{\rm val}, \kappa_{\rm sea}$).
The corresponding pion masses range from about 500 MeV to 1.1 GeV.
The results are extrapolated to the physical quark-mass point.
Our results indicate that 
there are two negative-parity $\Lambda$ states nearly degenerate 
at around 1.6 GeV, whereas no state as low as $\Lambda (1405)$ is observed.
By extracting the flavor components of each state, 
we find that the lowest (1st-excited) negative-parity state
is dominated by flavor-singlet (flavor-octet) component.
}

\begin{document}

\maketitle

\section{Introduction}

$\Lambda (1405)$ has been attracting much interest from several view points. 
$\Lambda (1405)$ is the lightest negative-parity baryon
in spite of the valence strange quark in it.
Among the $J^P = 1/2^-$ baryons, $\Lambda (1405)$ is much lighter than
the non-strange counterpart $N(1535)$, and isolated from the others.
It has no spin-orbit partner in the vicinity, 
taking into account that the lowest spin $3/2^-$ state is $\Lambda (1520)$.
Furthermore, the structure of $\Lambda (1405)$ remains mysterious. 
On one hand, $\Lambda (1405)$ is interpreted as a flavor-SU(3)-singlet 
three-quark state in conventional quark models.
On the other hand, 
$\Lambda (1405)$ could be interpreted as an antikaon-nucleon 
$\bar KN$ molecular bound state (B.E. $\sim$ 30 MeV). 
The binding energy of $\bar KN$ implies a
strong attraction between $\bar K$ and $N$
~\cite{Sakurai:1960ju,Dalitz:1967fp}, 
which may cause a new type of dense hadronic
matter, kaonic nuclei or kaonic nuclear matter
~\cite{Akaishi:2002bg,Yamazaki:2002uh,Akaishi:2005sn}.
We also expect that such $\bar KN$ bound states with large
binding energies can be regarded as compact 5-quark states.
The 5-quark picture of $\Lambda(1405)$ has advantages
that all five quarks can be placed in the lowest-lying $L=0$ state
forming a negative-parity baryon, and also 
that it requires no spin-orbit partner for $\Lambda (1405)$.

Therefore, the property of $\Lambda (1405)$ 
can be an important clue to new paradigm in hadron physics. 
We here study properties of $\Lambda(1405)$ using the lattice QCD formulation.
Though several lattice QCD studies
on $\Lambda (1405)$ have been performed so far
~\cite{Melnitchouk:2002eg,Nemoto:2003ft,Burch:2006cc,Ishii:2007ym},
most of them are based on quenched QCD, and 
few lattice QCD studies succeeded in reproducing the mass of $\Lambda (1405)$.
Moreover, in many cases, individual analyses with ``singlet'' and ``octet''
operators were performed,
where possible flavor mixings were not taken into account.
Little has been discussed on the lattice about the possible mixing
of flavor-SU(3)-octet and -singlet components induced by
the symmetry breaking.
It is then an intriguing issue to clarify the flavor structures
in excited-state hadron resonances.

Several possible reasons 
for the failure of reproducing $\Lambda (1405)$ in lattice QCD
were suggested through these studies;
lack of meson-baryon components due to quenching,
exotic (non-3-quark type) structure of $\Lambda (1405)$,
or insufficiency of the lattice volume in the simulations.
Resolving such difficulties requires 
unquenched lattice QCD calculation on a larger lattice volume
with varieties of interpolating operators.
Evaluating contaminations by scattering states
induced purely by sea quarks would be newly needed.
In this paper, we aim at clarifying the properties of $\Lambda (1405)$
with two-flavor full lattice QCD~\cite{Takahashi:2009bu},
employing the ``octet'' and ``singlet''
baryon operators to construct correlation matrices,
which enables us to extract the low-lying spectrum
as well as the mixing between octet and singlet
components in $\Lambda (1405)$.

\section{Lattice QCD setups}

\subsection{Simulation conditions}

We adopt the renormalization-group improved gauge action
and the ${\mathcal O}(a)$-improved quark action.
We adopt three different $\beta$'s,
$\beta = 1.80$, 1.95 and 2.10,
and corresponding lattice spacings are $a = 0.2150$, 0.1555 and 0.1076
fm~\cite{AliKhan:2001tx}.
We employ four different hopping parameters 
($\kappa_{\rm val}, \kappa_{\rm sea}$)
for each cutoff.
Corresponding pion masses range approximately from 500MeV to 1.1 GeV
at each $\beta$.
The details are found in Ref.~\citen{Takahashi:2009bu}

\subsection{Baryonic operators}

In order to extract the low-lying states in $S=-1$ and isosinglet channel,
we construct $2\times 2$ cross correlators
from the ``singlet'' and ``octet'' operators,
\begin{eqnarray}
&&{\eta}_{\bf 1}(x)
\equiv
\frac{1}{\sqrt{3}}
\epsilon^{abc} \nonumber \\
&&\times
\left(
u^a(x)[d^{T b}(x) C \gamma_5 s^c(x)] 
+
d^a(x)[s^{T b}(x) C \gamma_5 u^c(x)]
+
s^a(x)[u^{T b}(x) C \gamma_5 d^c(x)] 
\right) \nonumber
\label{defeta1}
\end{eqnarray}
\begin{eqnarray}
&&{\eta}_{\bf 8}(x)
\equiv
\frac{1}{\sqrt{6}}
\epsilon^{abc} \nonumber \\
&&\times
\left(
u^a(x)[d^{T b}(x) C \gamma_5 s^c(x)]
+
d^a(x)[s^{T b}(x) C \gamma_5 u^c(x)]
-2
s^a(x)[u^{T b}(x) C \gamma_5 d^c(x)]
\right) \nonumber
\label{defeta2}
\end{eqnarray}
It is easy to check that ${\eta}_{\bf 1}(x)$
(${\eta}_{\bf 8}(x)$) belongs to the singlet (octet) 
irreducible representation of the flavor SU(3).
We adopt point-type operators for the sink,
and extended operators, which are smeared
in a gauge-invariant manner, for the source.
Smearing parameters are chosen so that 
root-mean-square radius is approximately 0.5 fm.

\section{Lattice QCD Results}

\subsection{Hadronic masses}

\begin{figure}[hbt]
\begin{center}
\includegraphics[scale=0.205]{2448_pos.eps}
\includegraphics[scale=0.205]{2448_neg.eps}
\includegraphics[scale=0.27]{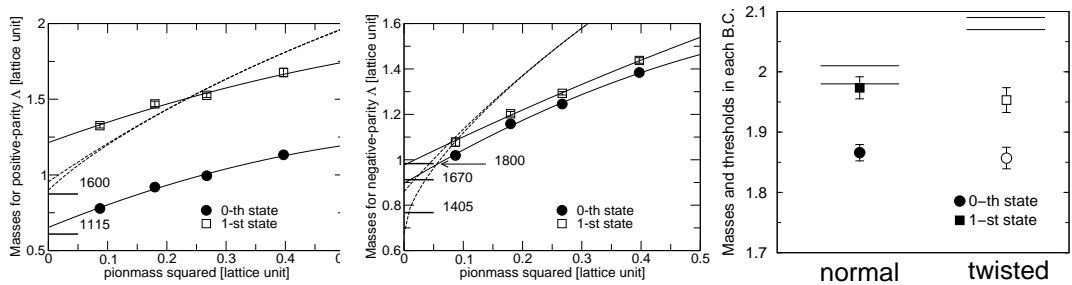}
\end{center}
\caption{\label{posmass}
[{\bf Left and middle panels}]
Masses of the lowest two $\Lambda$ states plotted as functions
of the squared pion mass,
which were obtained on $24^3\times 48$ lattice.
The filled circles (open squares) denote
the masses of lowest (1st-excited) state.
Two solid curves represent quadratic functions
used in the chiral extrapolation.
({\bf left}):
Two dashed lines indicate the $\pi\Sigma$ and the $\bar K N$ thresholds
in the presence of the relative momentum $p=\frac{2\pi}{L}$.
Two solid lines on the vertical axes show
the experimentally observed masses of $\Lambda(1115)$ and $\Lambda(1600)$.
({\bf middle}):
Two dashed lines indicate the $\pi\Sigma$ and the $\bar K N$ thresholds
with the relative momentum $p=0$.
Two solid lines on the vertical axes show
the experimentally observed masses of $\Lambda(1405)$ and $\Lambda(1670)$.
[{\bf Right panel}]
Lowest two eigen-energies in the $(J^P,S) = (1/2^-,-1)$ channel
under the normal (periodic) and the twisted boundary conditions.
The circles (squares) are for the ground (1st-excited) states.
The lower and upper solid lines respectively represent
the threshold energies of $\Sigma\pi$ and $N\bar K$ states
evaluated with normal/twisted boundary condition.
These data were obtained with the largest hopping parameter
(smallest quark mass),
and the corresponding pion mass is 550 MeV.
}
\end{figure}

Fig.~\ref{posmass} (left and middle panels) show
the eigen-energies in the positive- and negative-parity channels,
plotted as functions of the squared pion mass $m_\pi^2$,
which were obtained on $24^3\times 48$ lattice.
The lattice spacing is $a=0.1076$ fm and the lattice cut off
is $a^{-1}=1.83$ GeV.
Filled circles and open squares denote 
the energies of the ground states and the 1st-excited states, respectively.
The solid curves represent quadratic chiral fits 
as functions of squared pion mass $m_\pi^2$.
Solid lines at the vertical axes
indicate the empirical masses of $\Lambda(1115)$ and $\Lambda(1600)$
in the left panel (positive-parity states),
and those of $\Lambda(1405)$, $\Lambda(1670)$ and $\Lambda(1800)$
in the middle panel (negative-parity states).

The extrapolated values of the positive-parity ground state
agree very well with the mass of the ground-state $\Lambda(1115)$.
On the other hand, the 1st-excited state in this channel
lies much higher than $\Lambda(1600)$,
which is the 1st excited state experimentally observed so far.
The same tendency was reported in Ref.~\citen{Burch:2006cc}, and 
the situation is similar to the case of the Roper resonance, 
the non-strange SU(3) partner of $\Lambda (1600)$~\cite{Sasaki:2001nf}.

In the negative-parity channel (middle panel in Fig.\ref{posmass}),
the ground- and the 1st-excited states always have similar energies
at all the $\kappa$'s.
These eigen-energies have similar quark-mass dependences, and
the mass splittings are almost quark mass independent.
The chirally extrapolated values both lie around 
the mass of $\Lambda(1670)$ rather than $\Lambda(1405)$.
Similarly to the previous studies,
the mass of $\Lambda(1405)$ is not reproduced in our calculation.
While in quenched simulations in Refs.\citen{Melnitchouk:2002eg,Nemoto:2003ft}
such failure was regarded as an evidence
of possible meson-baryon molecule components in $\Lambda(1405)$,
our present {\it unquenched}
simulation contains effects of dynamical quarks
and thus should incorporate meson-baryon molecular states.

We here mention possible contaminations from scattering states.
Since our calculations contain dynamical up and down quarks,
scattering states could come into the spectra.
($N\bar K$ and $\Sigma\pi$ thresholds
are drawn in Fig.~\ref{posmass} as dashed lines.)
The energy of the 1st-excited state 
at the lightest u- and d-quark masses
is very close to the threshold on the lattice,
and could be contaminated by such scattering states.
There are two meson-baryon channels relevant in the present calculation,
$\pi\Sigma$ and $\bar K N$.
In the physical charge basis, we have 5 different thresholds
(2 for $\bar K N$, 3 for $\pi\Sigma$).
In order to distinguish resonance states from all these scattering states,
we impose the following boundary condition on the quark fields,
\begin{equation}
\psi(x+L)
=
e^
{\frac23\pi i}
\psi(x).
\end{equation}
Under such boundary condition for quark fields,
a hadronic state $\phi_{3k+n}(x)$ which contains $3k+n$ valence quarks obeys
\begin{equation}
\phi_{3k+n}(x+L)
=
e^
{\frac23n\pi i}
\phi_{3k+n}(x),
\end{equation}
and can have spatial momenta,
$
p_{\rm lat}
=
\frac{2\pi}{L}m+\frac{2n\pi}{3L}
\ \ 
(m \in {\rm Z})
$.

As a result, only states which consist of 3k valence quarks
can be zero-momentum states. Since other quark combinations 
inevitably have non-vanishing spatial momenta, their energies are raised up.
As long as we employ three-quark operators for baryon creation/annihilation,
hadronic states appearing in scattering states should contain
one or two valence quark(s), 
since sea-quark pairs themselves cannot carry flavors.

We plot in the right panel of Fig.~\ref{posmass} the lowest two eigen-energies
in the $(J^P,S) = (1/2^-,-1)$ channel
under the periodic and the twisted boundary conditions.
The open and filled circles (squares) are for the ground (1st-excited) states.
The lower and upper solid lines respectively represent
the threshold energies of $\Sigma\pi$ and $N\bar K$ states
under normal/twisted boundary condition.
These data were obtained with the largest hopping parameter
(smallest quark mass),
and the corresponding pion mass is 550 MeV.
The threshold energies are raised up
in the case of twisted boundary condition,
whereas the lowest two eigen-energies remain unchanged.
Thus we conclude that the observed states are insensitive to boundary conditions,
and contaminations from meson-baryon scattering states are small,
which would be due to the same reason as the absence of string breaking
in heavy-quark potentials from Wilson loops.

\subsection{Flavor structures}

The chiral unitary approach~\cite{Jido:2003cb} has suggested
that $\Lambda (1405)$ is not a single pole
but a superposition of two independent resonance poles.
This possible structure of $\Lambda (1405)$
is now attracting much interest,
and desired to be clarified in a model independent manner.
We investigate the flavor structures
of the ground and the 1st-excited states
obtained from the cross correlators of two operators.

In order to clarify the flavor structures of the low-lying states,
we evaluate the overlaps of the obtained states with the source and
sink interpolating fields.
We evaluate $g_0$ and $g_1$ defined as
\begin{eqnarray}
g_0^-
\equiv
\langle {\rm 0th} | \eta_{\bf 8} | {\rm vac}\rangle
/
\langle {\rm 0th} | \eta_{\bf 1} | {\rm vac}\rangle,
\ \ 
g_1^-
\equiv
\langle {\rm 1th} | \eta_{\bf 1} | {\rm vac}\rangle
/
\langle {\rm 1th} | \eta_{\bf 8} | {\rm vac}\rangle
\end{eqnarray}
Both $g_0^-$ and $g_1^-$ vanish when the SU(3)$_F$ symmetry is exact,
showing that the ground (1st excited) state is purely flavor singlet (octet) 
in the limit.
For the positive-parity states, we similarly define
\begin{eqnarray}
g_0^+
\equiv
\langle {\rm 0th} | \eta_{\bf 1} | {\rm vac}\rangle
/
\langle {\rm 0th} | \eta_{\bf 8} | {\rm vac}\rangle,
\ \ 
g_1^+
\equiv
\langle {\rm 1th} | \eta_{\bf 8} | {\rm vac}\rangle
/
\langle {\rm 1th} | \eta_{\bf 1} | {\rm vac}\rangle
\end{eqnarray}
In this case, we exchange the denominator and the numerator
since the flavor assignment in the SU(3)$_F$ limit of
the ground- and the 1st-excited states are opposite.

\begin{figure}[hbt]
\begin{center}
\includegraphics[scale=0.25]{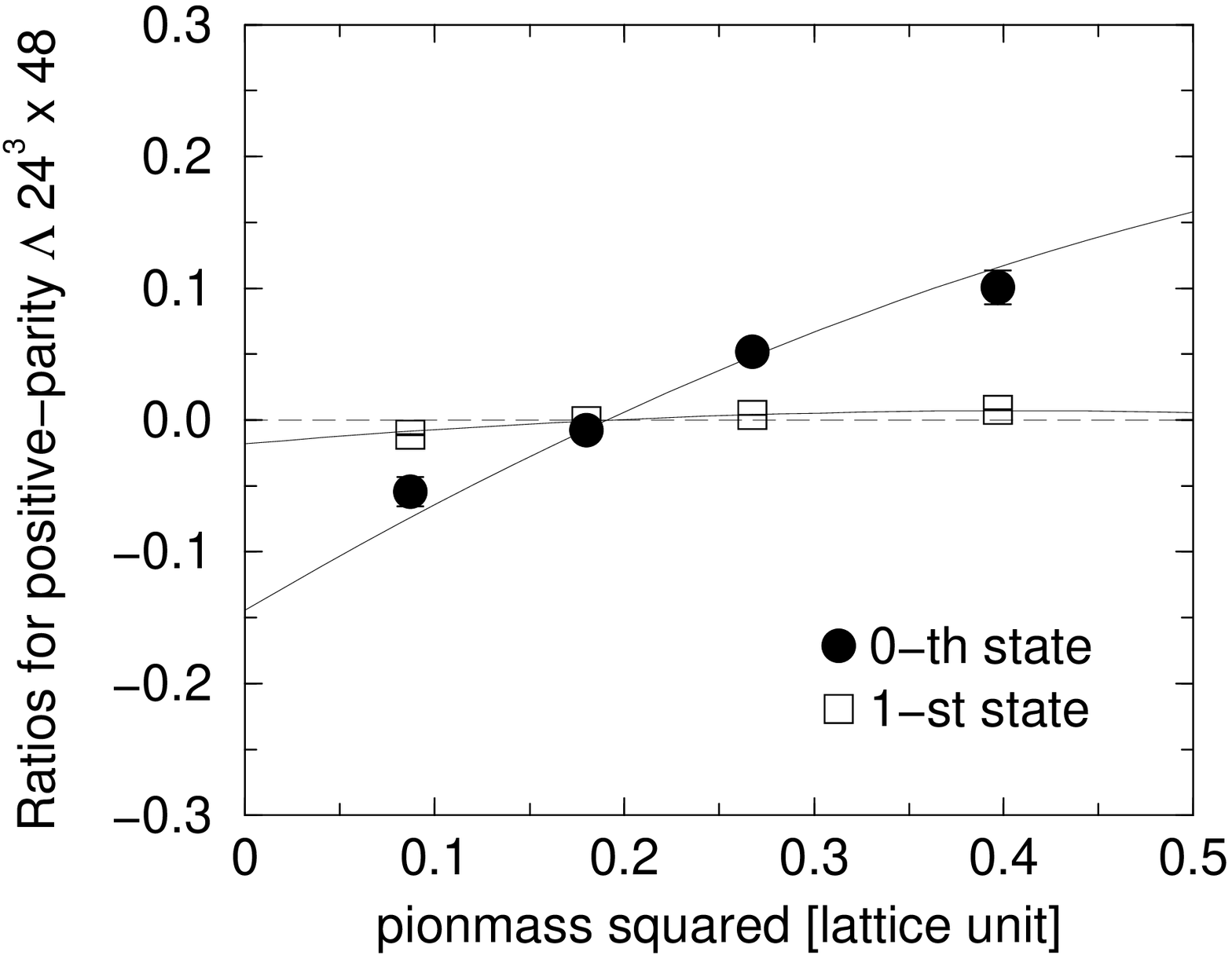}
\includegraphics[scale=0.25]{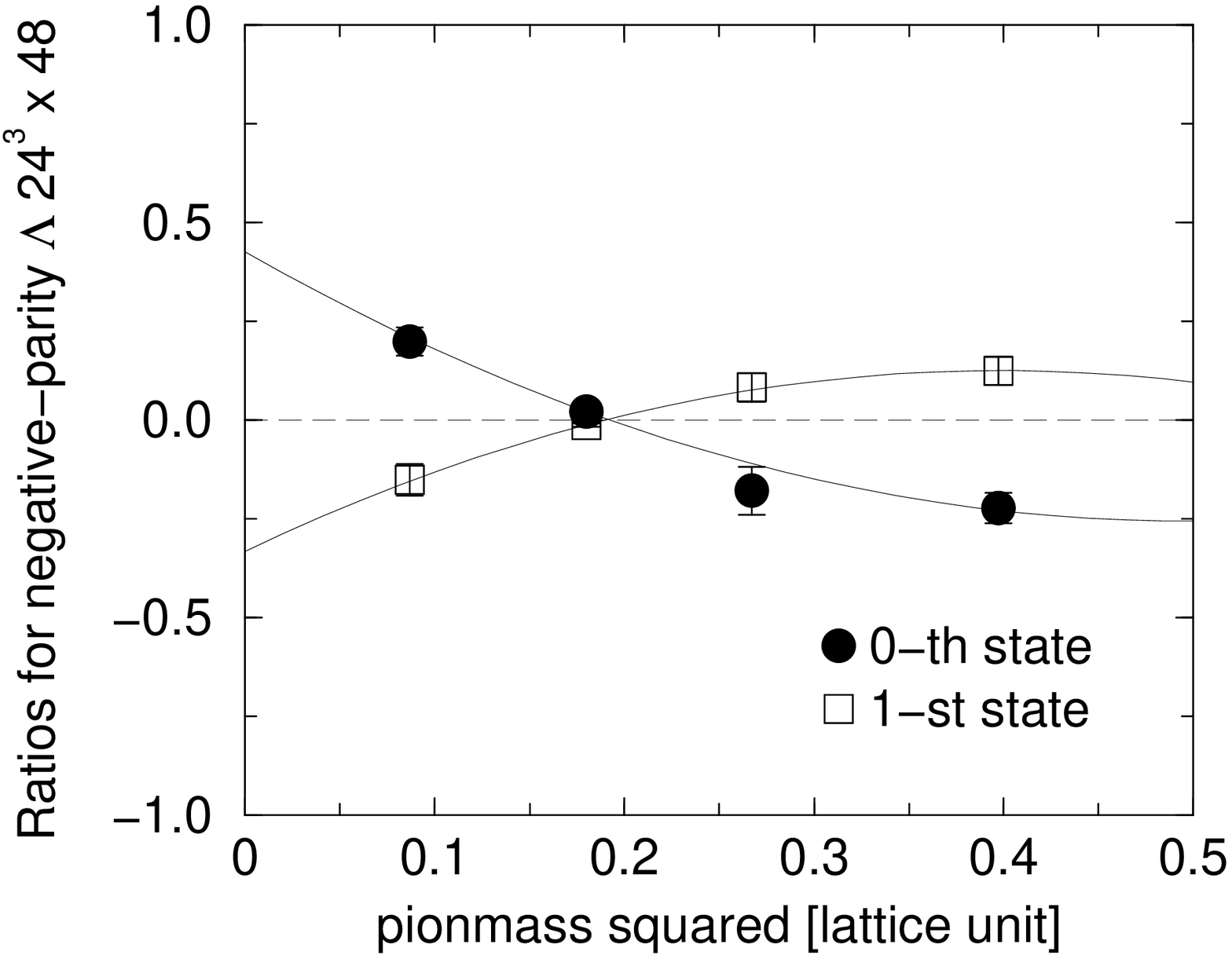}
\includegraphics[scale=0.25]{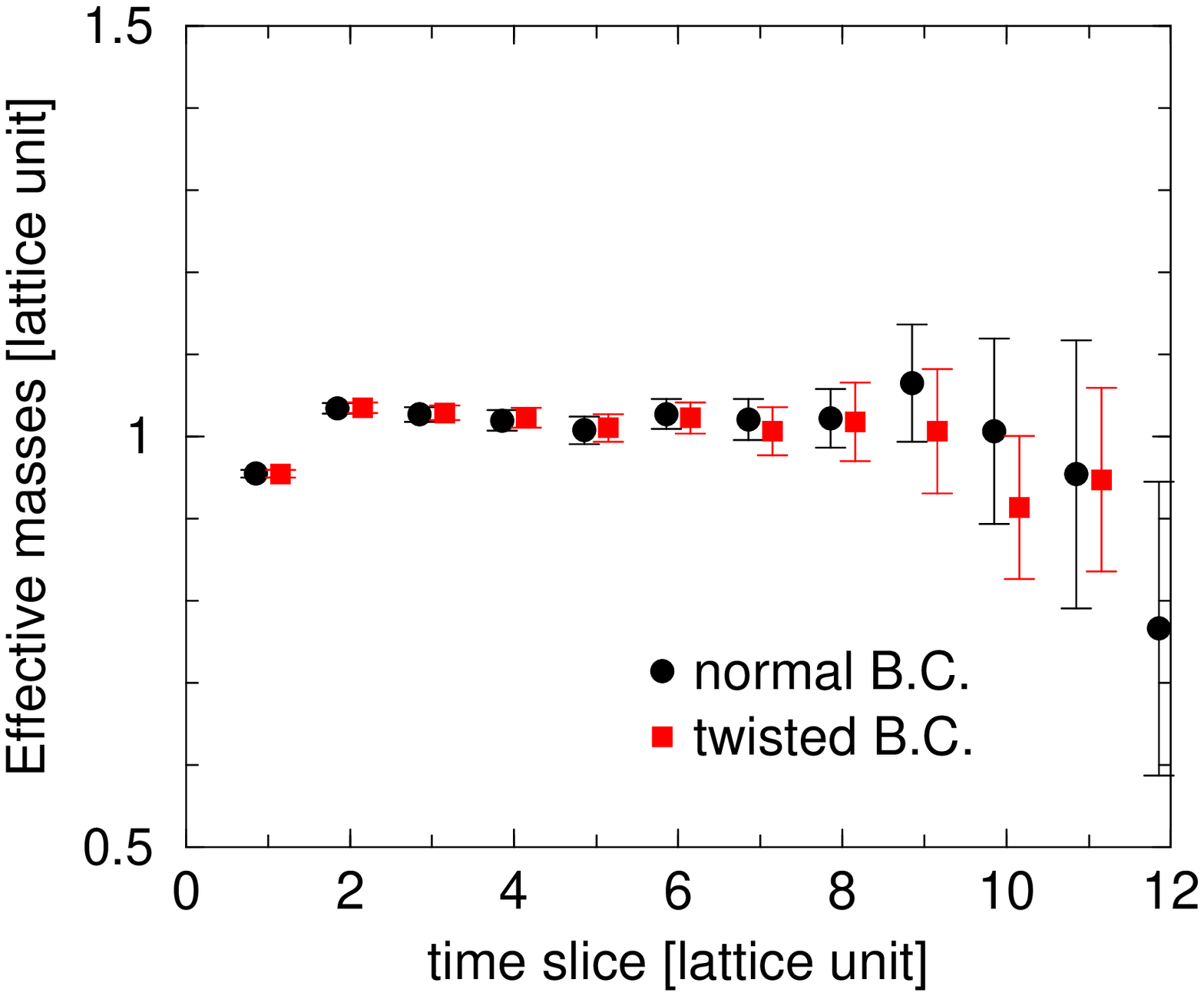}
\end{center}
\caption{\label{gpos}
[{\bf Left and middle panels}]
$g_i^\pm$ are plotted as a function of $m_\pi^2$.
Solid lines denote quadratic functions used in chiral extrapolation.
({\bf left}): The ratio $g_i^+$ in the positive-parity channel.
({\bf middle}): The ratio $g_i^-$ in the negative-parity channel.
[{\bf Right panel}]
Effective masses of the negative-parity $\Lambda$ ground state
under the normal (periodic) and the twisted boundary conditions.
}
\end{figure}

We show $g_0^\pm$ and $g_1^\pm$ as functions of pion-mass squared
for each lattice cutoff in Fig.~\ref{gpos} (Left and middle panels).
As is expected from the symmetry, such mixing coefficients
cross zero at the flavor symmetric limit
($\kappa_u=\kappa_d=\kappa_s$).
The data show smooth quark-mass dependence
toward the chiral limit in all the lattice-cut-off cases
and 
the dependences are almost lattice-cut-off independent.
We find that the magnitudes of operator mixing
get larger for larger flavor-symmetry breaking.
It is interesting that
the 1st-excited state in positive-parity channel 
is dominated by the flavor-singlet component
showing almost no contribution of octet components.

On the other hand, the mixing of the singlet and octet components
is generally weaker in the positive parity $\Lambda$'s.
It is natural because the splitting of the two states are large in the
SU(3)$_F$.
Nevertheless, it is interesting to see that the ground state, 
corresponding to $\Lambda(1115)$, contains significant (about 10\%) 
flavor singlet component in the chiral limit, which was not
expected from the simple quark model picture.

\section{Conclusion and Discussions}

An unquenched lattice QCD study for the low-lying $\Lambda$ baryon states
in the $S=-1$, $I=0$ and $J=1/2$ channel has been performed, focusing 
especially on the masses and flavor structures 
of the two lowest negative-parity $\Lambda$'s.
We have constructed $2 \times 2$ cross correlators 
from the ``octet'' and ``singlet'' baryonic interpolating fields,
and diagonalized them so that we can properly extract the information of 
the low-lying two states.
In our measurements, we have adopted the 2-flavor full-QCD gauge configurations
generated by CP-PACS collaboration with
the renormalization-group improved gauge action
and the ${\mathcal O}(a)$-improved quark action.
We have performed simulations at three different $\beta$'s,
$\beta = 1.80$, 1.95 and 2.10,
whose corresponding lattice spacings are $a = 0.2150$, 0.1555 and 0.1076
 fm, employing four different hopping parameters 
($\kappa_{\rm val}, \kappa_{\rm sea}$) for each cutoff, corresponding to 
the pion masses ranging from 500 MeV to 1.1 GeV.
The results have been extrapolated to the physical quark-mass region.

It is shown that the mass of the ground-state $\Lambda(1115)$ is reproduced 
well at all three $\beta$'s, while the 1st-excited positive-parity $\Lambda$ state 
lies much higher than the experimentally observed $\Lambda(1600)$.
The same tendency was reported in Ref.~\citen{Burch:2006cc}, and 
also in the case of  the Roper resonance, $N^*(1440)$, 
which is the non-strange SU(3) partner of $\Lambda (1600)$~\cite{Sasaki:2001nf}.
It should also be noted that no evidence of 
$\pi\Sigma$ or $\bar K N$ scattering states are seen in the calculation.

Our results for negative-parity $\Lambda$ states indicate that 
there are two $1/2^-$ states nearly degenerate 
at around $1.6 -1.7$GeV,
while no state as low as $\Lambda (1405)$ is observed.
We have revealed the flavor structures of these states
from the lattice data for the first time.
It is found that 
the lowest negative-parity $\Lambda$ state is dominated by flavor-singlet component.
The second state, which is less than 100 MeV above the ground state, 
is predominantly flavor octet.  
Thus we find that the first two negative-parity $\Lambda$'s have different flavor
structures.

However, they come significantly heavier than the experimentally observed
lowest-mass  state, $\Lambda(1405)$.
A possible straightforward interpretation is that these obtained states do
not correspond to the physical $\Lambda$(1405), but describe the higher
$\Lambda$ resonances.
In fact, the 2nd and 3rd negative-parity states lie at 1670 MeV
and 1800 MeV, both of which are three-star states in the Particle Data Group classification.
The present lattice-QCD data are consistent with these excited states. 
In the non-relativistic quark model approach, each of these states is classified as a flavor octet P-wave baryon.
We, however, have shown that the lower state is 
dominated by a flavor-singlet component.
The results here predict one flavor-singlet state 
and one flavor-octet state in the vicinity.
The reason for missing $\Lambda(1405)$ state will be 
its poor overlap with three-quark operators.
In fact, the inclusion of dynamical quarks does not
strongly enhance the signals of possible meson-baryon scattering states.
(See the later discussions.)
If $\Lambda (1405)$ is predominantly a meson-baryon molecular state,
such overlaps would be naturally small.

The other possibility is, of course, that
the lowest and the 2nd-lowest states describe physical 
$\Lambda(1405)$ and $\Lambda(1670)$
but the masses have been overestimated.
Considering that our simulation contains two-flavor dynamical quarks,
the failure of obtaining a light $\Lambda$ state could be attributed
either to
(1) uncertainty of the chiral extrapolation,
(2) strange-quark quenching,
(3) insufficient lattice volume or
(4) lack of chiral symmetry.

If $\Lambda (1405)$ is not a simple 3-quark state but 
predominantly a meson-baryon molecule state,
the $m_\pi$ dependence of its mass might be 
more complicated than simple polynomials,
as in the case of the $m_\pi$-dependence of the sigma-meson pole
in $\pi$-$\pi$ scattering~\cite{Hanhart:2008mx}.
The strange-quark quenching (2) 
seems to make the masses of octet baryons in positive
parity channel slightly ($\sim 10\%$) overestimated
in the present setups~\cite{AliKhan:2001tx}.
On the other hand, the deficiencies,(3) and (4), may cause the lowest
state not properly reproduced, supposing that the main component of
$\Lambda (1405)$ is a meson-baryon molecular state.
In order to check whether this conjecture is correct, simulations with
light dynamical quarks ($m_\pi \ll 500$ MeV)
and larger volume ($L \gg 2.5$ fm) will be required.

Upon the above conjecture,
we can further consider one interesting scenario that
these states both correspond to the physical $\Lambda(1405)$.
Then, the results may support its double pole structure proposed by the chiral unitary approach~\cite{Jido:2003cb}.
In our results, the lowest two states are almost degenerate at all the $\beta$'s (lattice spacing) and $\kappa$'s (quark masses).
Namely, the obtained two states are the signature of the double-pole resonance, but the
mass has not yet been reproduced because of the deficiency stated above.

It is also important to confirm that the meson-baryon scattering states
are not observed.
Despite that dynamical up and down quarks are included
and the meson-baryon thresholds appear around/below the obtained eigen-energies,
we have found no clear signal of the meson-baryon scattering states.
This fact is supported by the observation that the effective mass plots
under the normal and twisted boundary conditions
show no prominent difference.
We show in the right panel of Fig.\ref{gpos} the effective mass plots
of the ground state negative-parity $\Lambda$ under the normal and twisted b.c.'s.
They are obtained 
on the $24^3\times 48$ lattice with the largest hopping parameter.
Thus we conclude that
no scattering state appears in the present spectrum.
Also in Ref.~\citen{Bulava:2009jb},
in which excited-state nucleon spectrum was systematically and extensively 
investigated with two-flavors of dynamical quarks,
no clear signal of scattering states was found
and the importance of multi-quark operators was raised.

A similar situation can be found in the computation of the Wilson loops,
whose expectation values give us the potential
between a (heavy fundamental) quark and an antiquark.
In the presence of dynamical quarks,
such an interquark potential should saturate and flatten at some interquark distance,
where the confining string is broken and a quark-antiquark pair is created.
However, no one has ever observed successfully such a ``string breaking'' effect
by means of the Wilson loop~\cite{Heller:1994rz,Bolder:2000un}.
One possible reason for this phenomenon is that
the Wilson loop itself has poor overlaps with such broken strings.

In the $\Lambda$ spectrum, we may similarly conjecture 
that the scattering states or broad resonances,
which are sensitive to boundary conditions,
have little overlap with the 3-quark interpolating field operators.
More detailed investigation would be needed
for clarification of 
scattering states purely induced by dynamical quarks.


The flavor structures of the $\Lambda$ states
have been very well clarified using the variational method.
The octet and the singlet components
are mixed when the flavor-SU(3) symmetry is broken.
Actually, the ground (1st-excited) state 
is dominated by singlet (octet) component, and 
the contamination by another representation
is at most 20\% (5\% when squared) in our present analysis.
From these findings, we expect that the flavor-SU(3) symmetry
is not largely broken.
A similar conclusion was also derived for the study of the meson-baryon coupling 
constants in lattice QCD~\cite{Erkol:2008yj}.

Because the SU(3) breaking effect seems small, the analyses without SU(3) mixings adopted so far
~\cite{Melnitchouk:2002eg,Nemoto:2003ft,Burch:2006cc,Ishii:2007ym}
make sense to some extent.
The mixings, however, get larger towards the chiral limit
and
variational analyses would be essentially needed,
when we adopt much lighter quarks.
We have also found that the meson-baryon contents in each state
strongly depend on the meson-baryon couplings and their signs.
(See Ref.~\citen{Takahashi:2009bu}.)
Precise determination of the meson-baryon contents will require reliable determination of the couplings up to signs,
for which further lattice QCD calculations may be helpful.

\section*{Acknowledgements}
All the numerical calculations were performed on NEC SX-8R at CMC, Osaka university and BlueGene/L at KEK. The unquenched gauge configurations employed in our analysis were all generated by CP-PACS collaboration~\cite{AliKhan:2001tx}. This work was supported in part by the 21st Century COE `Center for Diversity and University in Physics'', Kyoto University and Yukawa International Program for Quark-Hadron Sciences (YIPQS), by the Japanese Society for the Promotion of Science under contract number P-06327 and by KAKENHI (17070002, 19540275, 20028006 and 21740181).

%

\end{document}